\documentclass[aps,prd,showpacs,preprintnumbers]{revtex4}
\usepackage{graphics}
\def \beq{\begin{equation}}
\def \eeq{\end{equation}}
\def \beqa{\begin{eqnarray}}
\def \eeqa{\end{eqnarray}}
\def \tr{{\rm Tr}\,}
\def \det{{\rm Det}\,}

\def \O{{\cal O}}
\def \ie{{\sl i.e.\/}}
\def \etal{{\sl et al.\/}}

\def \pl{{\sl Phys.\ Lett.\/}}
\def \pr{{\sl Phys.\ Rev.\/}}
\def \prl{{\sl Phys.\ Rev.\ Lett.\/}}

\begin{document}
 
\title{Simple patterns for non-linear susceptibilities near $T_c$}
\author{R.\ V.\ \surname{Gavai}}
\email{gavai@tifr.res.in}
\affiliation{Department of Theoretical Physics, Tata Institute of Fundamental
         Research,\\ Homi Bhabha Road, Mumbai 400005, India.}
\author{Sourendu \surname{Gupta}}
\email{sgupta@tifr.res.in}
\affiliation{Department of Theoretical Physics, Tata Institute of Fundamental
         Research,\\ Homi Bhabha Road, Mumbai 400005, India.}

\begin{abstract}
Non-linear susceptibilities upto the eighth order have been constructed in
QCD with 2 flavours of dynamical quarks. Beyond leading order, they exhibit
peaks at the cross over temperature, $T_c$. By analyzing their behaviour in
detail, we find that the dominant contributions near $T_c$ come from a set
of operators with a remarkably simple topology. Any effective theory of
QCD near $T_c$ must be able to explain these regularities.
\end{abstract}
\preprint{TIFR/TH/05-29, hep-lat/0507023}
\maketitle

Quark number susceptibilities (QNS) in QCD \cite{gott} are interesting
because they are measurable through event-to-event fluctuations of
conserved quantities in heavy-ion collisions \cite{ahm}.  Recent
determinations of the linear QNS in lattice QCD include those in
the continuum limit of the quenched theory \cite{valence}, the first
results in the high temperature phase of $N_f=2$ QCD
\cite{pushan,biswa,long} and the first computation in $N_f=2+1$ QCD
\cite{milc}. The non-linear susceptibilities (NLS) are a generalization
introduced in \cite{pressure,lat03} and have been used in finding
the Taylor expansion of the pressure of the QCD plasma at finite
chemical potential. The linear combinations used for pressure were
also reported in $N_f=2$ QCD \cite{biswa}.

Here we report on systematic simplicities of these quantities that
we discovered in our investigation of QCD with light dynamical quarks.
These simple patterns which we find here for the first time may be
consistent with weak coupling theory in the high temperature phase
of QCD. However, in the vicinity of the finite temperature cross over at
$T_c$, we find a different simple pattern.  It seems
possible to incorporate it into a simple model of the physics of
the cross over. A few of these results have been discussed in
\cite{long}. Here we complete the study of the NLS started there.

\begin{figure}
\begin{center}
   \scalebox{0.5}{\includegraphics{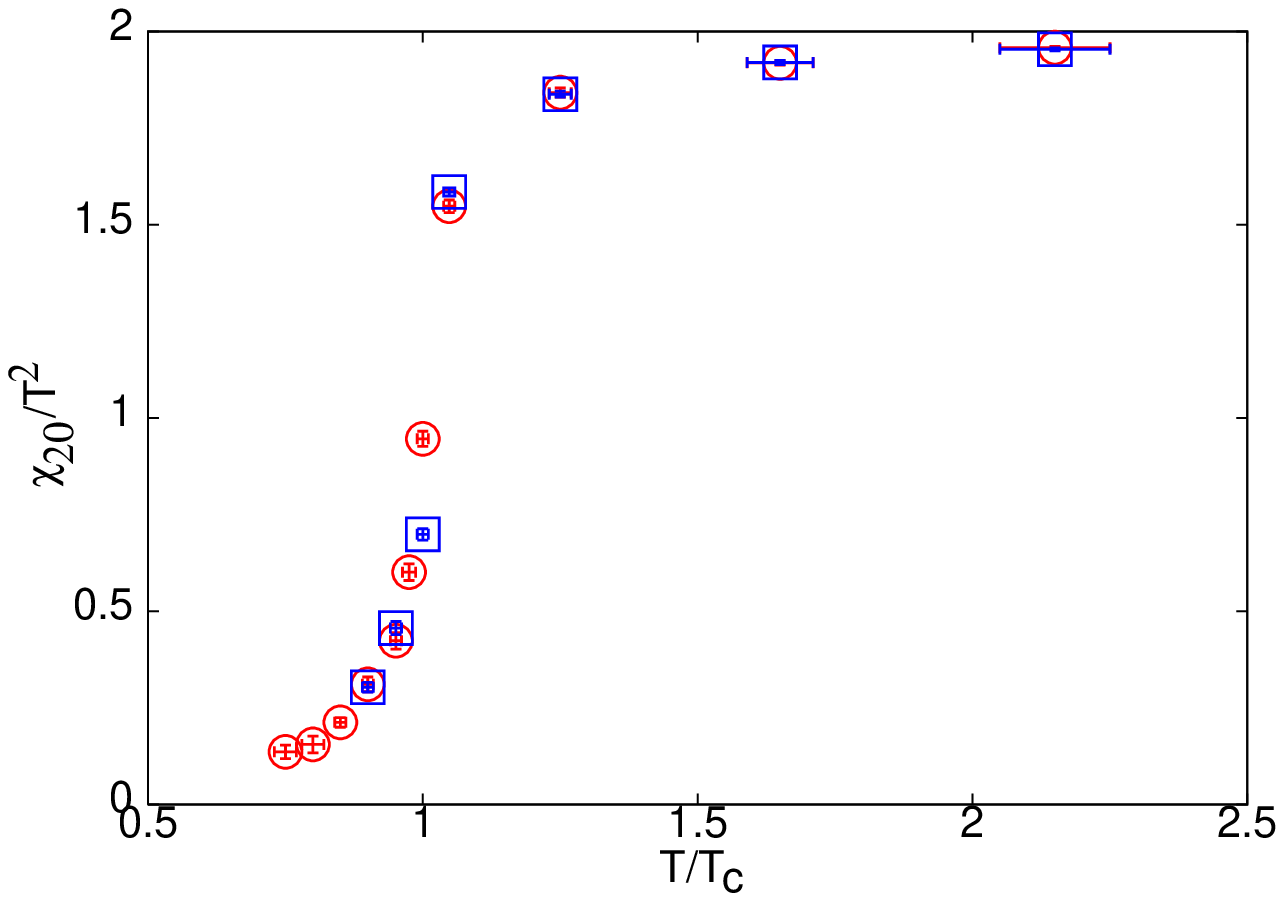}}
   \scalebox{0.5}{\includegraphics{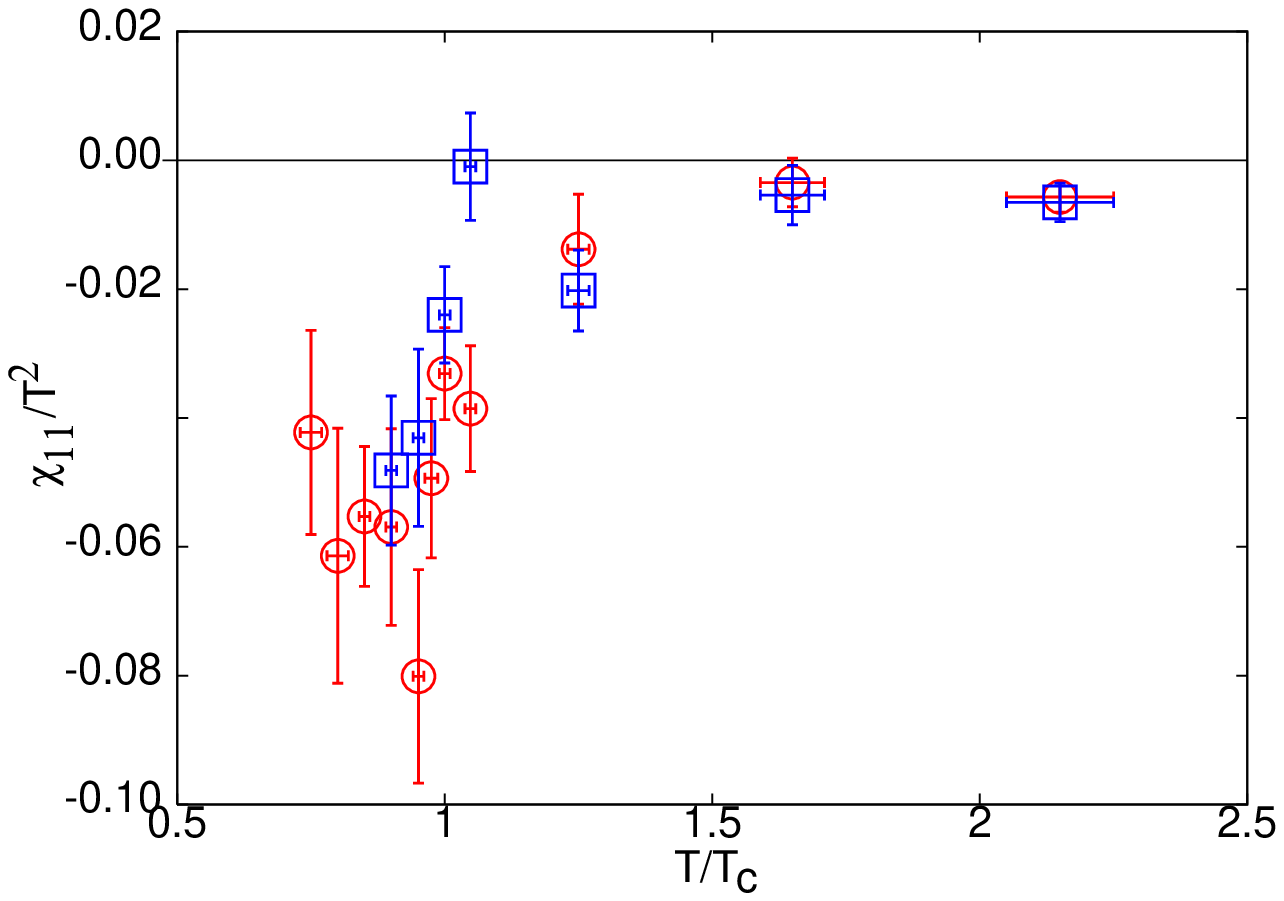}}
\end{center}
\caption{$\chi_{20}/T^2$ varies smoothly across $T_c$, and behaves roughly as an
   order parameter, being small in the hadronic phase and large in the plasma.
   $\chi_{11}/T^2$ is small in the hadronic phase, perhaps peaks near $T_c$ and
   is not significant in the plasma phase. Data from lattice sizes $4\times16^3$
   (circles) and $4\times24^3$ (boxes) are shown.}
\label{fg.qns}\end{figure}

\begin{figure}
\begin{center}
   \scalebox{0.65}{\includegraphics{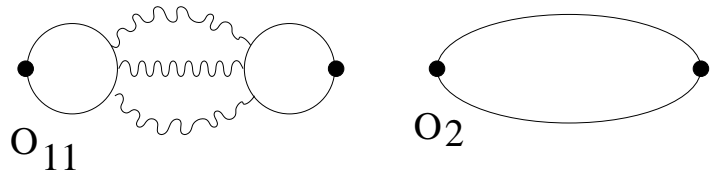}}
\end{center}
\caption{The operator $\O_{11}$ is shown on the left with the smallest number
   of gluon connections between the two fermion loops. The contribution is
   naively of order $g^6$, but a computation shows that it is actually
   $g^5\log g$ \cite{bir}. The operator $\O_2$ shown on the right is fermion
   line connected and hence the leading contribution shown is of order 1. The
   black dots denote insertions of $\gamma_0$ arising from the derivatives with
   respect to the chemical potential.}
\label{fg.dgm2}\end{figure}

The partition function for QCD at temperature $T$ and chemical potentials
$\mu_f$ for each of $N_f$ flavours, can be written in the form
\beq
   Z(T,\{\mu_f\}) = \int{\cal D}U\,{\rm e}^{-S_G(T)}\,\prod_f\det M_f(m_f,T,\mu_f),
\label{part}\eeq
where $S_G$ is the gluon part of the action and $M$ denotes the Dirac
operator.  The pressure,
\beq
   P(T,\{\mu_f\})=-\,\frac FV = \left(\frac TV\right)\log Z(T,\{\mu_f\}),
\label{pres}\eeq
which is a convex function of $T$ and $\mu_f$,
can be expanded in a Taylor series about the point where all the
$\mu_f=0$ \cite{pressure}. 

In this paper we examine staggered fermions with $N_f=2$ and a small but 
non-vanishing quark mass, $m_u=m_d=m$. The $N$-th order derivatives in the 
Taylor expansion then can be taken $n_u$ times with respect to $\mu_u$ and
$n_d=N-n_u$ times with respect to $\mu_d$. This is the non-linear
quark number susceptibility (NLS), which we write as $\chi_{n_u,n_d}$.
This new notation streamlines a more cumbersome notation which
was used earlier. The translation table between these two notations
can be understood from the relations---
\beq
   \chi_{20}\equiv\chi_{uu}=\chi_{dd}\equiv\chi_{02},\quad\chi_{11}\equiv\chi_{ud},\quad
   \chi_{40}\equiv\chi_{uuuu}=\chi_{dddd}\equiv\chi_{04},\quad\chi_{22}\equiv\chi_{uudd},
   \quad etc,
\label{trans}\eeq
where we have used flavour symmetry to write $\chi_{n_u,n_d}=\chi_{n_d,n_u}$.

The Taylor expansion of the pressure can be written as
\beq
   \Delta P(T,\mu_u,\mu_d) \equiv
   P(T,\mu_u,\mu_d)-P(T,0,0) = \sum_{n_u,n_d} \chi_{n_u,n_d}\;
        \frac{\mu_u^{n_u}}{n_u!}\, \frac{\mu_d^{n_d}}{n_d!}.
\label{presst}\eeq
The NLS above can be written down in terms of the derivatives of
$Z$. From the expression in eq. (\ref{part}) it is clear that the derivatives
with respect to the $\mu_f$ land entirely on the determinants. Now,
since $\det M=\exp{\tr\log M}$, the first derivative gives $(\det
M)'=\tr(M^{-1} M')\det M\equiv\O_1\det M$. Higher derivatives can
be found systematically using the additional relation $M M^{-1}=1$,
which yields $(M^{-1})'=-M^{-1} M' M^{-1}$. Our notation for operators
is that $\O_n'=\O_{n+1}$, and $\O_{lmn\cdots}=\O_l\O_m\O_n\cdots$.
The expectation values $\langle\O_{2n+1}(\mu_f=0)\rangle=0$ by CP symmetry.
The derivatives of $Z$ can be written in terms of expectation values
of certain operators involving powers of traces of products of
inverses and derivatives of the Dirac operator. Diagrammatic methods
for their evaluation were developed in \cite{pressure,zakopane} and
explicit expressions were written down in \cite{long}.

We report on results obtained using the configurations generated
in the study reported in \cite{long}. Details of our simulations
and statistics can be found there. These results have been obtained on
lattices with temporal extent $N_t=4$, and varying $N_s$, with the
spatial volume being large. The quark mass has been fixed in physical
units to be such that $m_\pi/m_\rho=0.31\pm0.01$, about 50\% larger
than in the real world, 
making this the smallest quark mass at which NLS have been studied.
Details of how the temperature scale is set
on the lattice can also be found in \cite{long}. In physical units
we find that the cross over temperature, $T_c$ is $m_\rho/T_c=5.4\pm0.2$.
%The quark mass is such that at $T=0$ one has the ratio $m_\pi/m_\rho=0.3$,
%making this the smallest quark mass at which NLS have been studied.

The volume dependence of $T_c$ has been remarked upon in \cite{long};
we see evidence of some volume dependence in the bare coupling at
the cross over, but the scale has larger uncertainties, so a finite
size scaling study of the shift of $T_c$ with $V$ performed at these
lattice cutoffs $a$ will not be very useful. However, strong finite
volume effects on the NLS were found when the spatial lattice extent
was too small, $N_s<4N_t$. In the remainder of this study, therefore,
we concetrate on the NLS obtained with $N_s=16$, using data obtained
with $N_s=24$ to make cross checks of the results. At $T_c$, the
finite volume shift in the results is significant, but become
negligible on moving slightly away--- to $0.95T_c$ or $1.05T_c$,
for example.

The two leading terms in the series, the diagonal QNS, $\chi_{20}$,
and the off-diagonal QNS, $\chi_{11}$, have been computed before.
For completeness we display results from \cite{long} in Figure
\ref{fg.qns}.  Note that $\chi_{11}=(T/V)\langle\O_{11}\rangle$,
which is a quark-line disconnected diagram. Also, one can see that
$\chi_{20}-\chi_{11}=(T/V)\O_2$, which is quark-line connected.
Diagrammatic representations of these are shown in Figure \ref{fg.dgm2}.
Recall that for $T>T_c$, these diagrams have been computed in
weak coupling theory, giving reasonable agreement with the lattice
results \cite{bir,alexi,mustafa}.

A counting rule for the minimum number of gluon lines needed in a
quark-line disconnected diagram was obtained in \cite{bir} by noting
that effectively the diagrams are Abelian, and Furry's theorem
holds, \ie, the number of $\gamma_\mu$ insertions must be even.
Among these must be counted the insertions of $\gamma_0$ arising
from taking derivatives with respect to the chemical potential. For
$\O_{11}$ one gluon exchange is ruled out for reasons of gauge
invariance, two by the counting rule, and hence three gluons are
needed, as shown in Figure \ref{fg.dgm2}.

In figure \ref{fg.qns} some volume dependence is visible in the
immediate vicinity of $T_c$. The high temperature behaviour of
$\chi_{20}/T^2$ is consistent with our earlier results in \cite{pushan},
and, therefore, is compatible with the predictions of \cite{bir,alexi}.
The results on $\chi_{11}/T^2$ are also completely compatible with
earlier results in \cite{pushan} after correcting for a division
by an extra factor of $(T/V)$ for $\chi_{11}/T^2$ reported there.
Comparison with the recent results of \cite{biswa,milc} are harder
to perform since the actions and quark masses are different.

At the fourth order, there are five operators--- $\O_4$, $O_{31}$,
$\O_{22}$, $\O_{112}$ and $\O_{1111}$. The last four are quark-line
disconnected operators. The connected parts of the operators enter
into the expressions for the NLS \cite{pressure,long}. In this
paper, we decompose the NLS into connected parts of these operators,
such as $(T/V) \langle \O_{22} \rangle_c$.  Since comparisons are
always with connected parts, we indulge in slight notation-abuse
by dropping the subscript often. We remind the reader of the
definitions of the connected parts at the fourth order---
\beqa
\nonumber &&
   \biggl\langle\O_{1111}\biggr\rangle_c=
   \left[\biggl\langle\O_{1111}\biggr\rangle
        -3\biggl\langle\O_{11}\biggr\rangle^2\right],
\\ \nonumber &&
   \biggl\langle\O_{112}\biggr\rangle_c=
   \left[\biggl\langle\O_{112}\biggr\rangle
        -\biggl\langle\O_{11}\biggr\rangle\biggl\langle\O_2\biggr\rangle\right],
\\  &&
   \biggl\langle\O_{22}\biggr\rangle_c=
   \left[\biggl\langle\O_{22}\biggr\rangle -\biggl\langle\O_2\biggr\rangle^2\right].
\label{nf4inv}\eeqa
$\biggl\langle\O_{31}\biggr\rangle$ and $\biggl\langle\O_4\biggr\rangle$
are connected pieces by themselves; the former by virtue of the
fact that $\langle\O_n\rangle=0$ for odd $n$, the latter because
it is the largest loop at this order.  In \cite{pressure,long} we
have shown that each distinct operator topology is a physical
observable in a version of QCD with appropriate number of quark
flavours.

\begin{figure}
\begin{center}
\scalebox{0.65}{\includegraphics{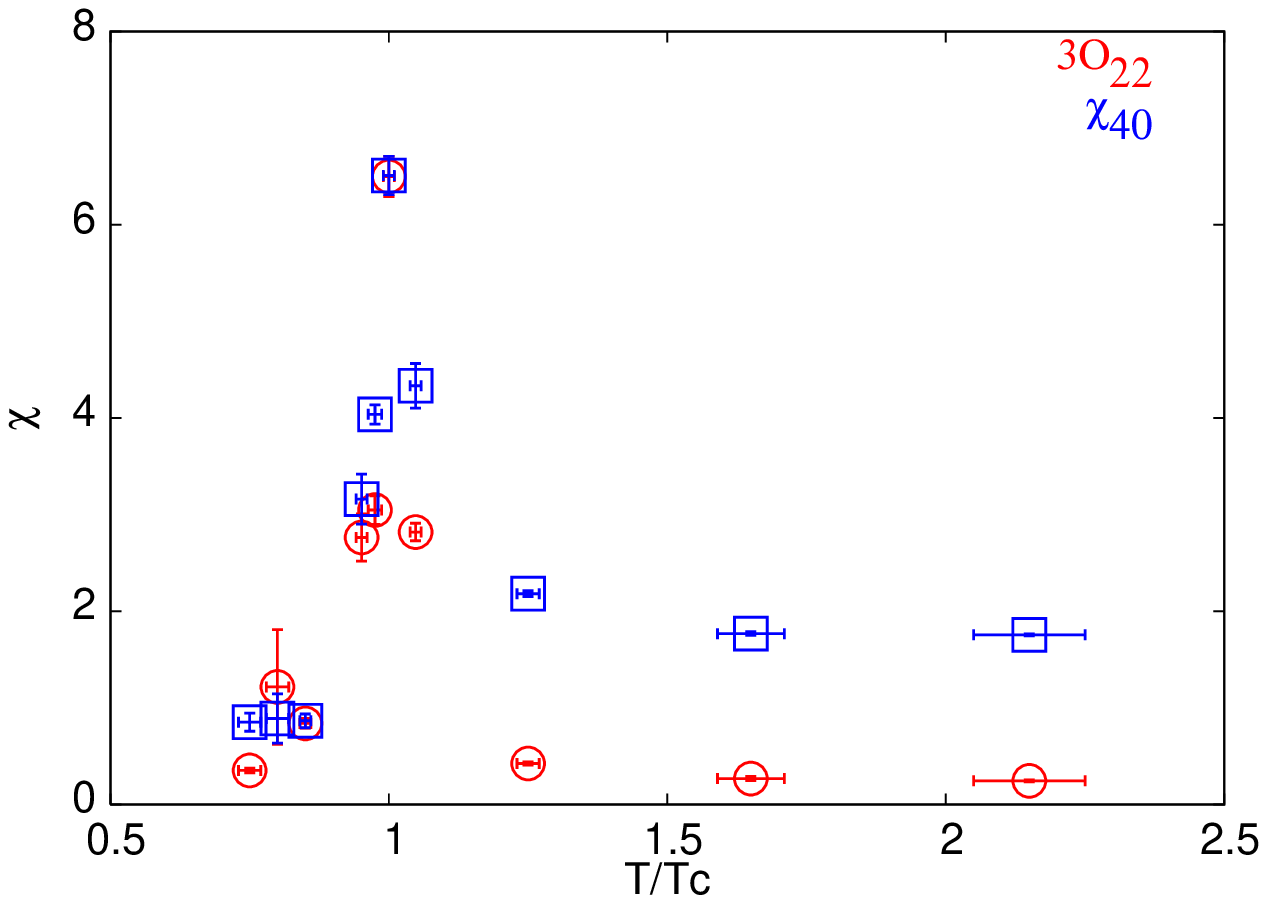}}
\scalebox{0.65}{\includegraphics{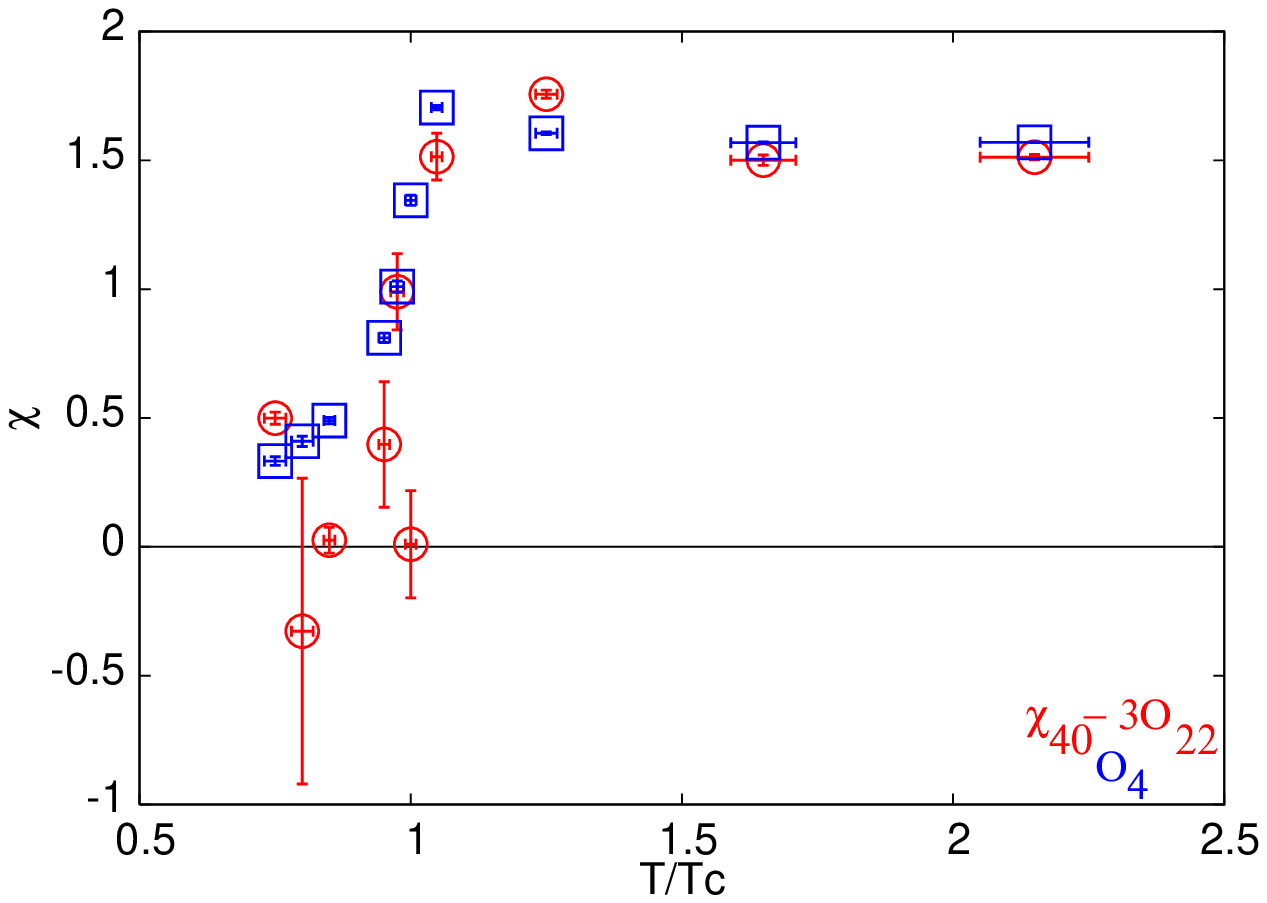}}
\end{center}
\caption{$\chi_{40}$ (boxes) peaks at $T_c$, and the peak is entirely due
   to the term in $(T/V)\langle\O_{22}\rangle$ (circles), as shown on the left.
   After subtracting this out, one gets a much smoother function (circles
   in the right panel), which agrees well with $(T/V)\langle\O_4\rangle$ 
   (boxes).  Data are from lattice sizes $4\times16^3$.}
\label{fg.ord4}\end{figure}

\begin{figure}
\begin{center}
   \scalebox{0.65}{\includegraphics{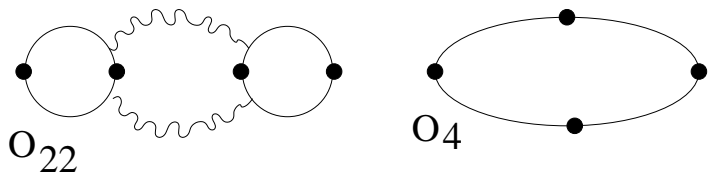}}
\end{center}
\caption{One of the contributions to the operator $\O_{22}$ is shown on the
   left with the smallest number of gluon connections between the two fermion
   loops allowed by the counting rules of \cite{bir}. Other contributions
   correspond to permuting the gluon lines and operator insertions along each
   quark line, while keeping the loop topology fixed. The operator $\O_4$
   shown on the right is fermion line connected and hence of order 1. These
   diagrams are expected to give accurate results in the plasma phase.}
\label{fg.dgm4}\end{figure}

We show our results for the QNS $\chi_{40}$ in Figure \ref{fg.ord4},
where we also plot the connected part of $(T/V) \langle \O_{22}\rangle$
multiplied by the coefficient with which it enters into $\chi_{40}$.
This operator appears to take care of the peak in the QNS near
$T_c$. In Figure \ref{fg.ord4} we have also shown the difference
between these two quantities. The peak disappears and the remainder,
in the high temperature phase, is saturated by $(T/V) \langle \O_4
\rangle$. Like $\O_2$, this expectation value is also like an order
parameter, being small in the low $T$ phase, and large on the other
side of $T_c$.

In this range of temperatures, the two major contributions to the
fourth order QNS are from $\O_4$ and $\O_{22}$. We also find this
kind of peak in $\chi_{22}$, where it again matched the peak in
$(T/V) \langle \O_{22}\rangle$. No other QNS at this order has
contribution from this operator, and also show little sign of a
comparable peak near $T_c$.  $\O_4$ gives no contribution to any
other QNS, and, compatible with this, we see that all other 4th
order QNS are very small above $T_c$. A similar behaviour is also
seen on the $4\times24^3$ lattice.

It is interesting that the counting rules of \cite{bir} show that the
two largest contributions above $T_c$ should come from precisely these
operators. $\O_4$ is of order 1, and the connected part of $\O_{22}$
shown in Figure \ref{fg.dgm4} is naively of order $g^4$. In comparison,
$\O_{31}$ is of order $g^6$, $\O_{112}$ is of order $g^8$ and $\O_{1111}$
is of order $g^{12}$. These naive powers may be modified into some logarithms
in the computation.

\begin{figure}
\begin{center}
\scalebox{1.00}{\includegraphics{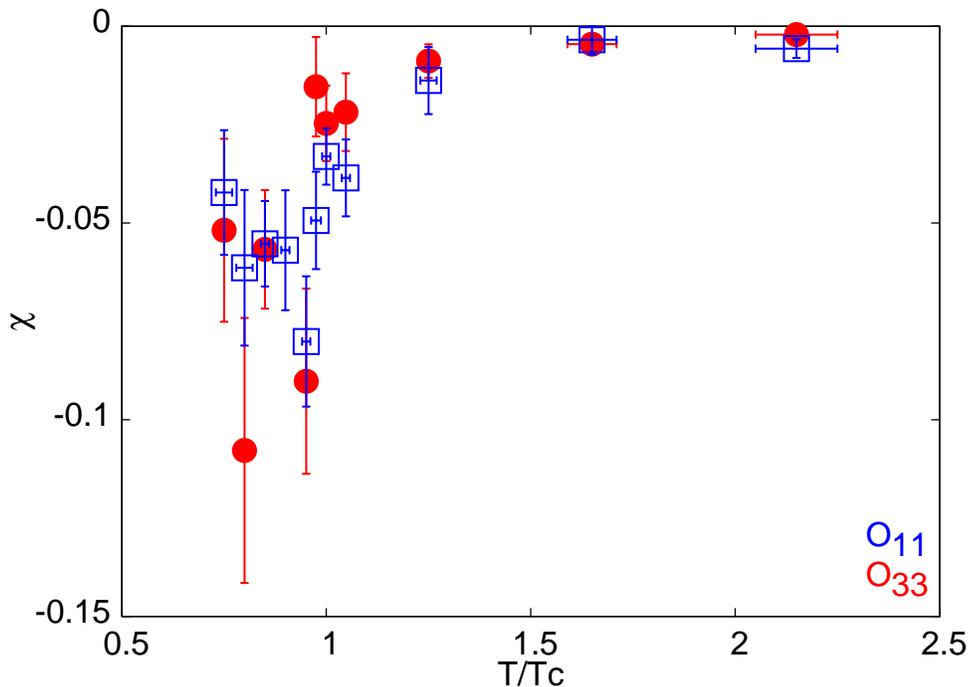}}
\end{center}
\caption{$\chi_{11}$ (boxes) and $(T/V) \langle \O_{33}\rangle$ (circles)
   obtained on a $4\times16^3$ lattice.}
\label{fg.c33}\end{figure}

At the sixth order we have eleven topologically distinct operators
$\O_6$, $\O_{51}$, $\O_{42}$, $\O_{33}$, $\O_{114}$, $\O_{123}$,
$\O_{222}$, $\O_{1113}$, $\O_{1122}$, $\O_{11112}$ and $\O_{111111}$.
The determination of the NLS are also significantly more expensive than
the linear QNS, requiring many more vectors in the stoachstic evaluation
of the traces \cite{long}. One result is that the measurements are more
noisy at higher orders. Nevertheless, it is possible to make significant
statements about the structure of these operators.

One interesting point, illustrated in Figure \ref{fg.c33}, is the
qualitative similarity between $\chi_{11}$ and
$(T/V)\langle\O_{33}\rangle$. Both are small in the high $T$ phase,
possibly peak in the vicinity of $T_c$, and are comparable to other
operators in the low $T$ phase. We have previously argued that the
increase in the ratio $\chi_{11}/\chi_{20}$ with decreasing $T$
implies that the fermion sign problem becomes more severe, thus
restricting the usefulness of all the recent methods which have
been developed to handle this problem. The observation in Figure
\ref{fg.c33} extends this argument to finite chemical potential.

\begin{figure}
\begin{center}
\scalebox{0.65}{\includegraphics{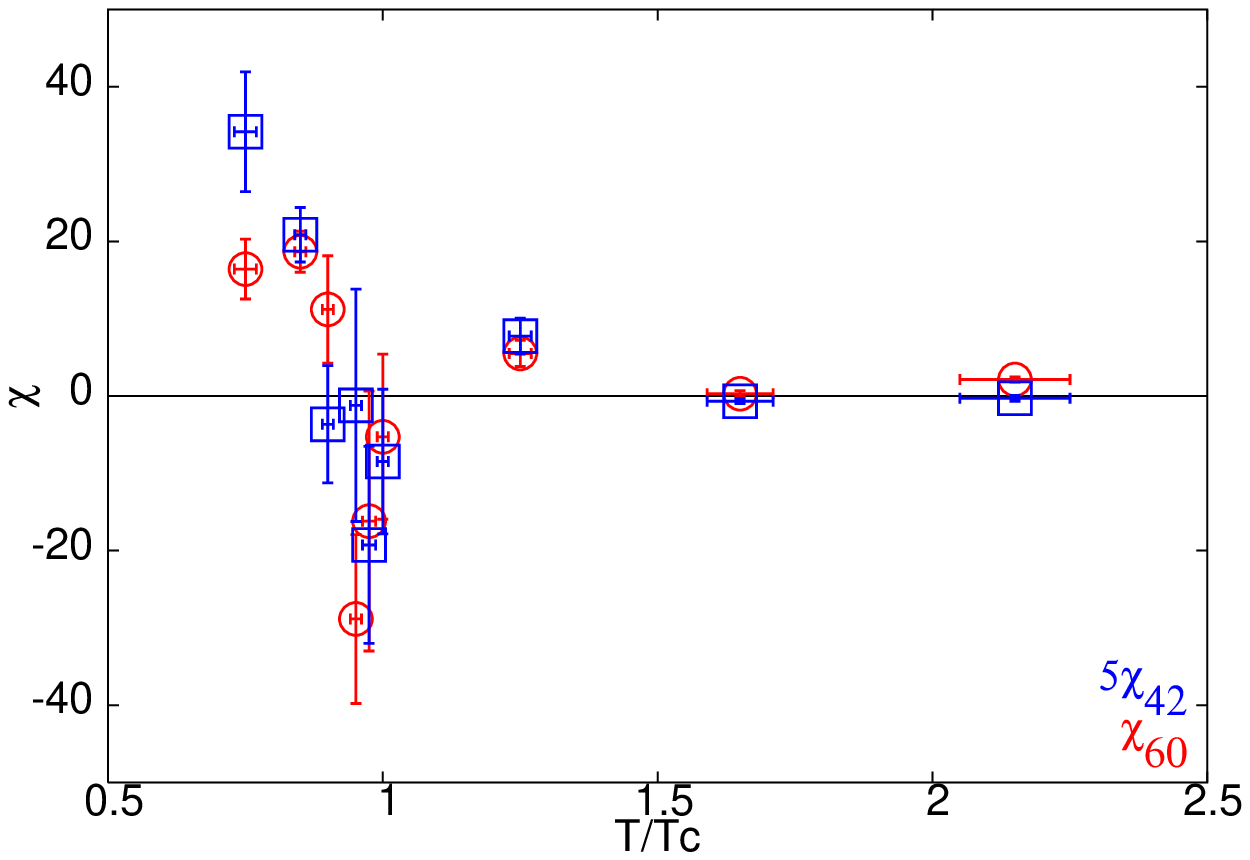}}
\scalebox{0.65}{\includegraphics{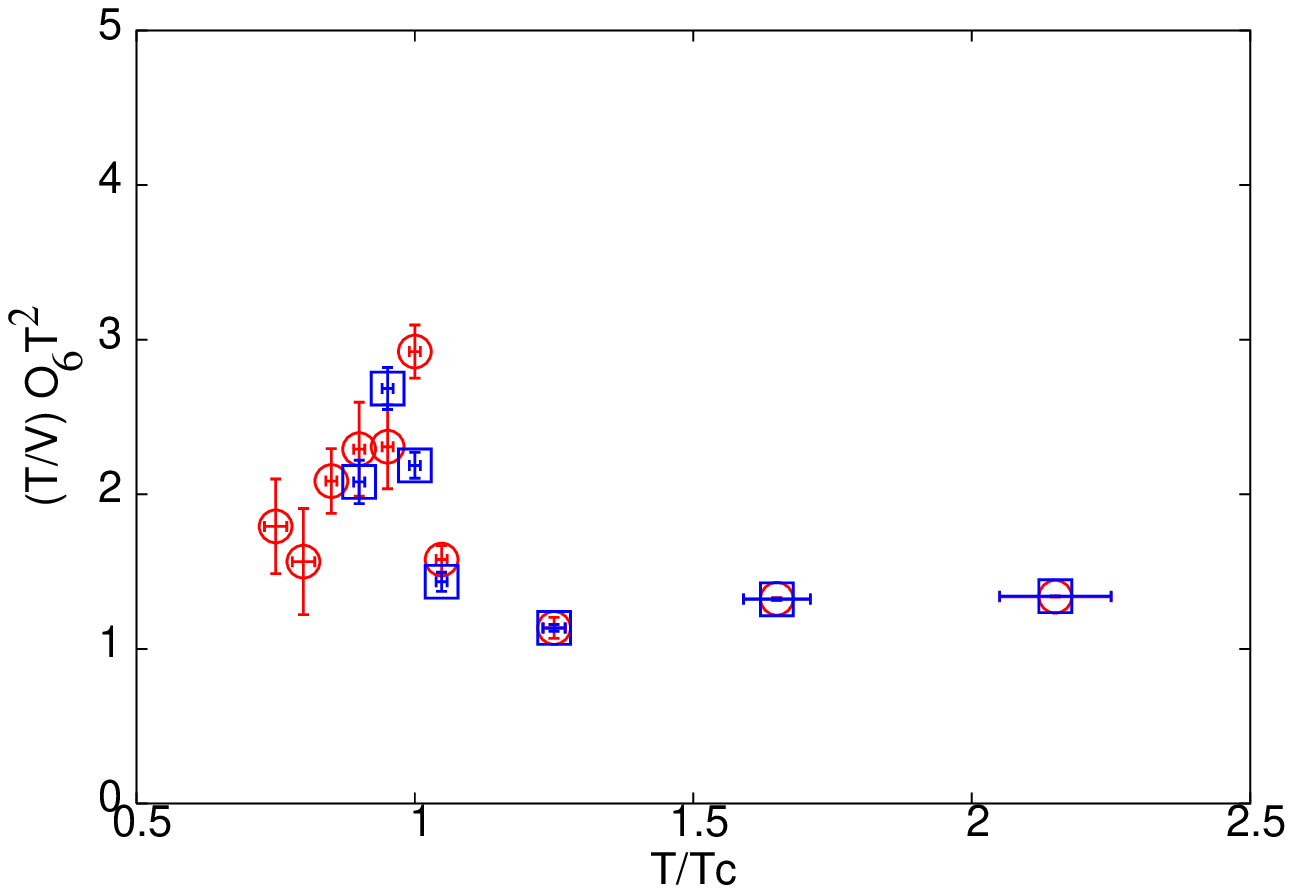}}
\end{center}
\caption{In the first panel we show $\chi_{60}$ (circles) and $5\chi_{42}$
   (boxes) as found on a $4\times16^3$ lattice. The two are normalized such
   that they have equal contribution from $\O_{222}$. The second
   panel shows $(T/V)\langle\O_6\rangle T^2$ on $4\times16^3$
   (circles) and $4\times24^3$ (boxes) lattices. Note the difference
   in the scales of the two figures.}
\label{fg.nls6}\end{figure}

\begin{figure}
\begin{center}
\scalebox{0.65}{\includegraphics{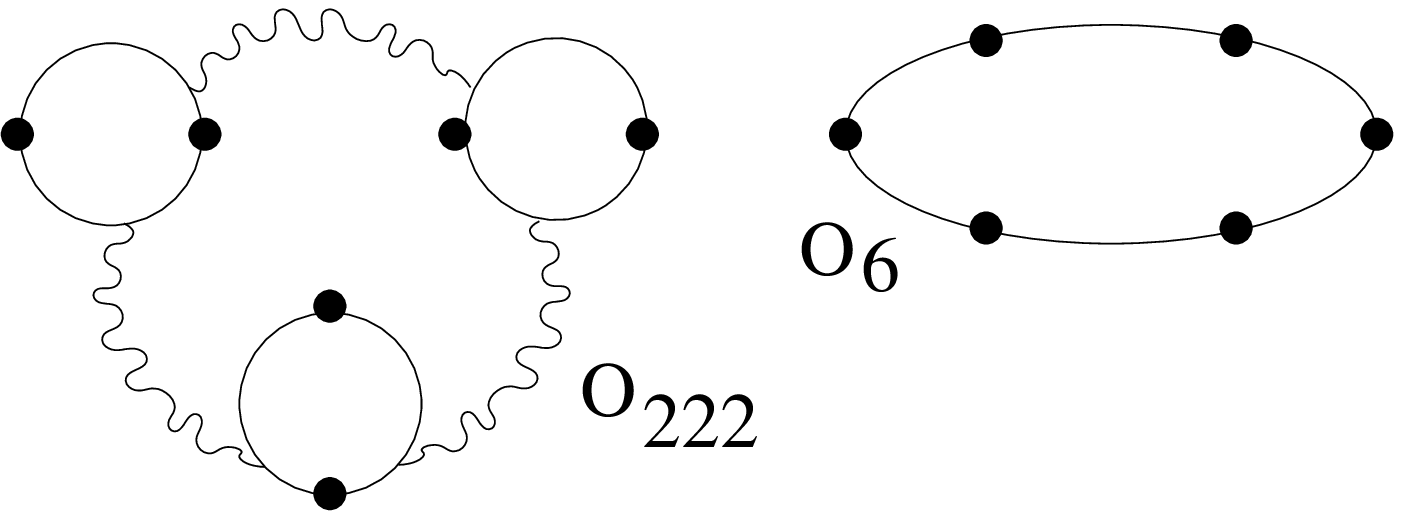}}
\end{center}
\caption{One of the contributions to the operator $\O_{222}$ is shown on the
   left with the smallest number of gluon connections between the two fermion
   loops allowed by the counting rules of \cite{bir}. Other contributions
   correspond to permuting the gluon lines and operator insertions along each
   quark line, while keeping the loop topology fixed. The operator $\O_6$
   shown on the right is fermion line connected and hence of order 1. These
   diagrams are expected to give accurate results in the plasma phase.}
\label{fg.dgm6}\end{figure}

\begin{figure}
\begin{center}
\scalebox{0.65}{\includegraphics{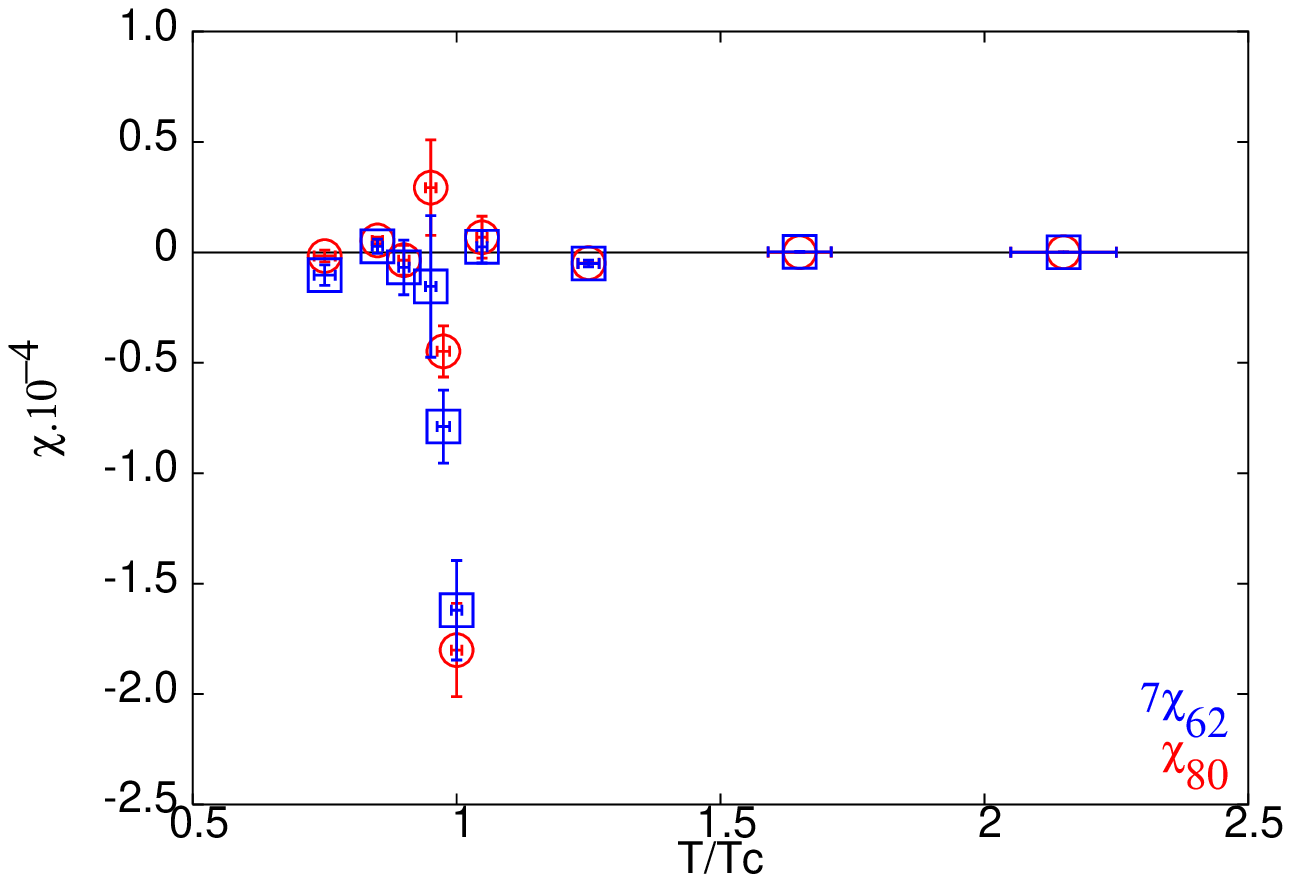}}
\scalebox{0.65}{\includegraphics{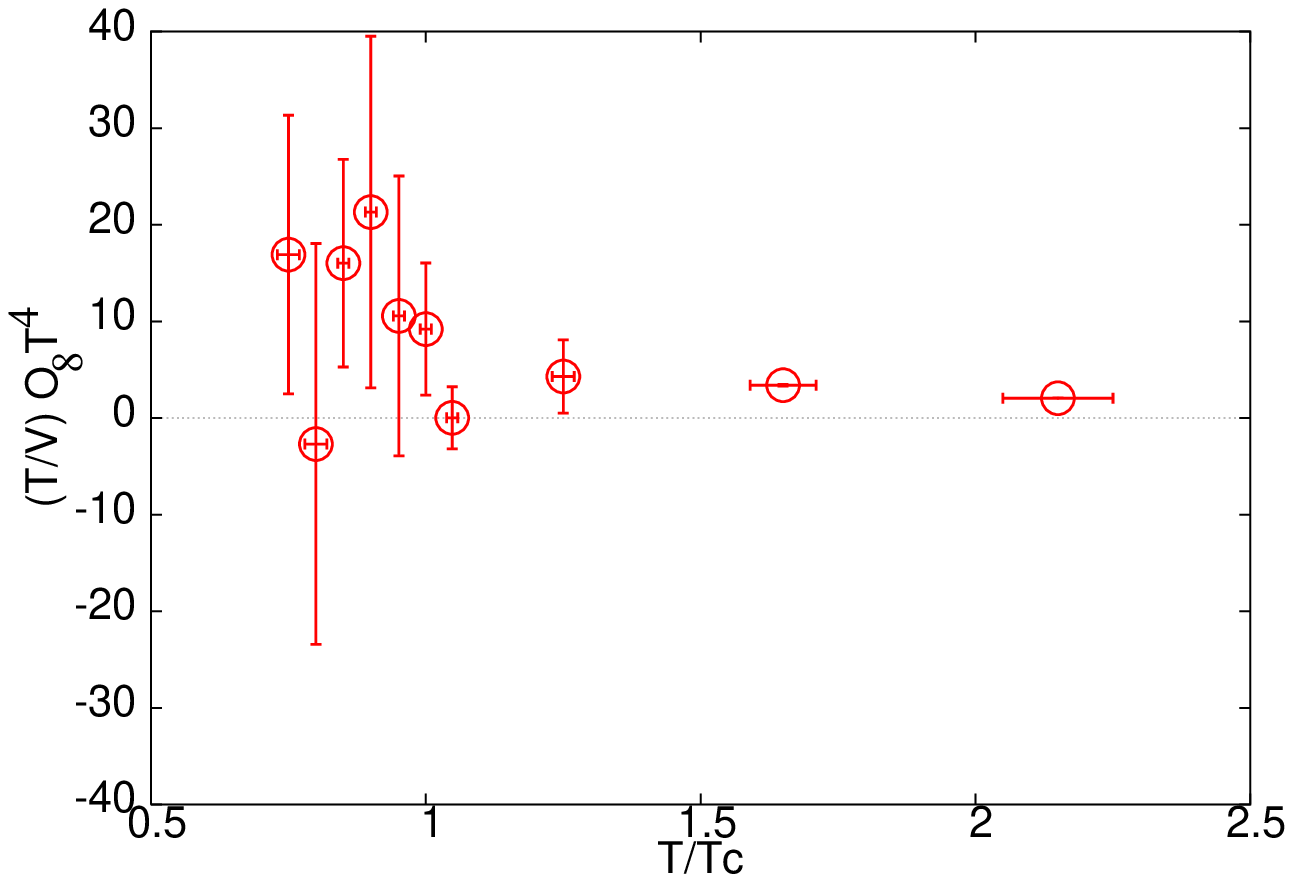}}
\end{center}
\caption{In the first panel we show $\chi_{80}$ (circles) and $7\chi_{62}$
   (boxes) as found on a $4\times16^3$ lattice. The two are normalized such
   that they have equal contribution from $\O_{2222}$. The second
   panel shows $(T/V)\langle\O_8\rangle T^4$ on a $4\times16^3$
   lattice. Note the difference
%  (circles) and $4\times24^3$ (boxes) lattices. Note the difference
   in the scales of the two figures.}
\label{fg.nls8}\end{figure}

\begin{figure}
\begin{center}
\scalebox{1.00}{\includegraphics{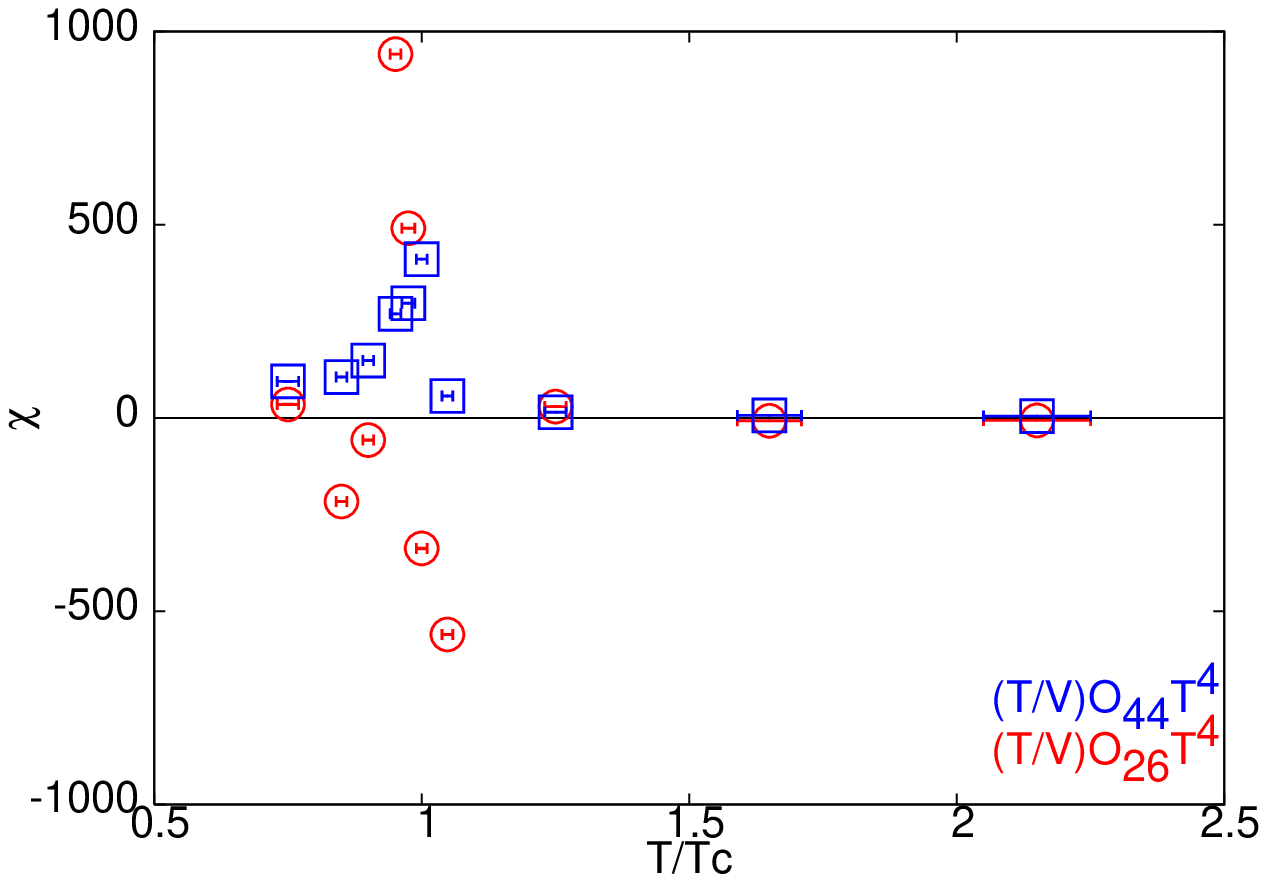}}
\end{center}
\caption{The connected parts of the expectation values of $\O_{26}$ (circles)
   and $\O_{44}$ (boxes) as found on a $4\times16^3$ lattice. The expectation
   values are normalized by $T/V$ and rendered dimensionless through
   a multiplication by $T^4$.}
\label{fg.offd8}\end{figure}

However, at higher temperatures, such contributions are small. The
dependence of $\chi_{60}$ on $T$ is shown in Figure \ref{fg.nls6}.
The peak at $T_c$ is due to contributions from $\O_{222}$, as we
demonstrate by plotting along with this the values of $\chi_{42}$
normalized so that the two have equal contribution from $\O_{222}$.
The difference is small; for $T>T_c$ it is saturated by $\O_6$,
which is much smaller than the peak, but much larger than $\O_{222}$.
The operator $\O_{24}$ also peaks at $T_c$, but the value at the
peak is negligible in comparison with $\O_{222}$. Power counting
shows that $\O_6$ is of order 1, $\O_{24}$ is of order $g^4$, but
$\O_{222}$ is of order $g^6$. The form of the operators is shown in
Figure \ref{fg.dgm6}. This is the lowest order at which we
first find explicitly that the perturbative power counting of the high
temperature phase does not extend down to $T_c$.

This pattern recurs at the eighth order, as we display in Figure
\ref{fg.nls8}.  There is a peak in some of the susceptibilities at
$T_c$, but this can be ascribed to $\O_{2222}$. The high temperature
phase is dominated by a non-vanishing value of $\O_8$, which is
much lower than the peak.  Other operators at the eighth order which
may peak at $T_c$ are $\O_{26}$ and $\O_{44}$. As we illustrate in
Figure \ref{fg.offd8}, they indeed have interesting behaviour near
$T_c$. However, these operators are numerically negligible compared
to the value of $\O_{2222}$. In the high temperature phase the power
counting rules show that $\O_8$ is of order 1, $\O_{26}$ and $\O_{44}$
are of order $g^4$, whereas $\O_{2222}$ is of order $g^8$. The pattern
of dominance near $T_c$ therefore has nothing to do with power counting
in $g$.

In summary then, we have found a very pleasing pattern for the NLS.
In the hadronic phase, all operators seem to have comparable
expectation values.  This is not unexpected. In the hadronic vacuum,
at $T=0$, many different operators have vacuum expectation values,
which are all typically expected to be of similar order. Above
$T_c$, we have an extremely simple pattern, in which the NLS are
dominated by the operators with a single quark loop, $\O_n$, and
the expectation values $(T/V)\langle\O_n\rangle T^{n-4}$ are all
in the range of 1--2. This pattern seems to be organized by
weak-coupling power counting arguments, but it would be useful to
have precise estimates of these operators through perturbative
computations.

It follows from this observation, that the pressure at finite
chemical potential has contributions from all even terms, but the
numerical importance of the terms decreases factorially at high
temperature. As shown in \cite{analy}, in a free field theory at
finite $\mu$, the pressure can be separated into a quark piece and
an antiquark piece, each of which has contributions to all even
orders in $\mu$, which cancel to give a pressure which contains
only terms upto order $\mu^4$.  These small terms in the pressure
can be thought of as little shift in these pieces caused by a weak
coupling, such that the cancellation becomes incomplete.  Such a
mismatch between particle and antiparticle is possible because a
chemical potential explicitly breaks CP invariance.

\begin{figure}
\begin{center}
\scalebox{1.00}{\includegraphics{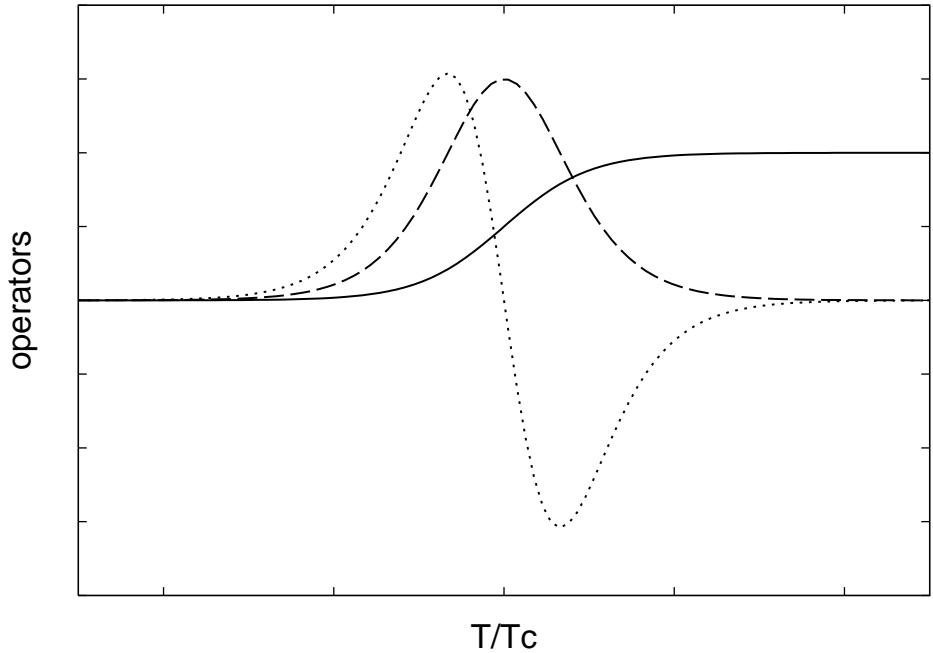}}
\end{center}
\caption{The expectations for the NLS near $T_c$ based on an effective
   theory of QCD near the phase transition in which the composite operator
   $\O_2$ is identified as the order parameter.}
\label{fg.form}\end{figure}

The most unexpected regularity that we have found is in the vicinity
of $T_c$. Here, the NLS are dominated by a composite operator which
is made up of appropriate numbers of fermion loops with two $\gamma_0$
insertions in each, \ie, with an appropriate number of $\O_2$.  Our
observation suggests that it may be possible to write down effective
long-distance theories in which this composite bosonic operator is
treated as a field operator whose expectation value shows the correct
cross over behaviour. In that case $\langle\O_{22}\rangle_c$ would
be the susceptibility of this field, and being proportional to the
temperature derivative of $\langle\O_2\rangle$, would peak, as
observed. The expectation value $\langle\O_{222}\rangle_c$ would
be proportional to the next derivative of $\langle\O_2\rangle$.
Then the $T$-dependence of these quantities at $\mu_f=0$ would have
the shapes shown Figure \ref{fg.form}.

In an effective 3-d spatial Landau theory of the form that we
suggest, $\O_2$ can be taken to be a two point function built from
one polarization of a vector operator. Under the symmetries of the
transfer matrix that builds the equilibrium correlation functions,
\ie, the screening correlators, this polarization mixes with the
scalar \cite{irreps}. It has been suggested that the scalar crucially
impacts the physics of the phase transition in the chiral limit
\cite{nakamura}, because of the fact that it becomes massless at
that point. This is the situation in the chiral limit; it would be
interesting to see predictions from such models for the behaviour
of these NLS at finite quark mass.

{\bf Acknowledgements:}
This computation was carried out on the Indian Lattice Gauge Theory
Initiative's CRAY X1 at the Tata Institute of Fundamental Research.
It is a pleasure to thank Ajay Salve for his administrative support
on the Cray.

\end{document}